\documentclass[
5p, twocolumn,
number, sort&compress
]{elsarticle}

\makeatletter
\def\ps@pprintTitle{%
 \let\@oddhead\@empty
 \let\@evenhead\@empty
 \def\@oddfoot{\hfill\today}%
 \let\@evenfoot\@oddfoot}
\makeatother

\usepackage{mathptmx}
\usepackage{amsmath}
\usepackage{amssymb}
\usepackage{graphicx}
\usepackage{xspace}
\usepackage{textcomp}
\usepackage[switch]{lineno}

\newcommand*\patchAmsMathEnvironmentForLineno[1]{%
  \expandafter\let\csname old#1\expandafter\endcsname\csname #1\endcsname
  \expandafter\let\csname oldend#1\expandafter\endcsname\csname end#1\endcsname
  \renewenvironment{#1}%
     {\linenomath\csname old#1\endcsname}%
     {\csname oldend#1\endcsname\endlinenomath}}%
\newcommand*\patchBothAmsMathEnvironmentsForLineno[1]{%
  \patchAmsMathEnvironmentForLineno{#1}%
  \patchAmsMathEnvironmentForLineno{#1*}}%
\AtBeginDocument{%
\patchBothAmsMathEnvironmentsForLineno{equation}%
\patchBothAmsMathEnvironmentsForLineno{align}%
\patchBothAmsMathEnvironmentsForLineno{flalign}%
\patchBothAmsMathEnvironmentsForLineno{alignat}%
\patchBothAmsMathEnvironmentsForLineno{gather}%
\patchBothAmsMathEnvironmentsForLineno{multline}%
}

\begin{document}


\title{Seeing Structural Evolution of Organic Molecular Nano-crystallites \\ Using 4D Scanning Confocal Electron Diffraction}

\author[fau]{Mingjian Wu\corref{cor1}}
\ead{mingjian.wu@fau.de}

\author[fau]{Christina Harrei\ss}

\author[ncem]{Colin Ophus}

\author[fau]{Erdmann Spiecker\corref{cor2}}
\ead{erdmann.spiecker@fau.de}

\cortext[cor1]{Corresponding author}
\cortext[cor2]{Corresponding author}

\address[fau]{Institute of Micro- and Nanostructure Research \& Center for Nanoanalysis and Electron Microscopy (CENEM), \\ Department of Materials Science, Universit\"at Erlangen-N\"urnberg, Cauerstra{\ss}e 3, D-91058 Erlangen, Germany.}
\address[ncem]{National Center for Electron Microscopy, Molecular Foundry, Lawrence Berkeley National Laboratory, 1 Cyclotron Road, Berkeley, CA, USA, 94720}

\begin{abstract}
Direct observation of organic molecular nanocrystals and their evolution using electron microscopy is extremely challenging, due to their radiation sensitivity and complex structure. 
Here, we introduce 4D-scanning confocal electron diffraction (4D-SCED), which enables direct {\em in situ} observation of bulk heterojunction (BHJ) thin films. 
4D-SCED combines confocal electron microscopy with a pixelated detector to record focused spot-like diffraction patterns with high angular resolution, using an order of magnitude lower dose than previous methods. 
We apply it to study an active layer in organic solar cells, namely DRCN5T:PC$_{71}$BM BHJ thin films. 
Structural details of DRCN5T nano-crystallites oriented both in- and out-of-plane are imaged at ~5 nm resolution and dose budget of \texttildelow $5~e^-/\AA^2$. 
We use {\em in situ} annealing to observe the growth of the donor crystals, evolution of the crystal orientation, and progressive enrichment of PC$_{71}$BM at interfaces. 
This highly dose-efficient method opens new possibilities for studying beam sensitive soft materials.
\end{abstract}

\maketitle


The properties of organic semiconductors and device performance, particularly in bulk heterojunction (BHJ) organic solar cells, is dictated by the nano-crystalline structure and morphology. 
This is due to the high anisotropy of opto-electronic properties of the constituent molecules or polymers, and their directional assembly into (semi-)crystals. 
The orientation relationship between molecule, crystal and morphological features, interface character of donor/acceptor components, degree of net-work percolation of the nano-scaled carrier transport channels in BHJ therefore govern the device performance, which all evolve sensitively depending on the processing conditions \cite{Kassar2016, Min2017, Min2016, Sun2014, Wang2015a}. 
However, revealing the nanoscale structures at high spatial resolution using electron microscopy methods is challenged by radiation sensitivity of these soft materials \cite{Chen2020, Egerton2019} and the complexity of their structures. 
Diffraction imaging, also called four dimensional-scanning transmission electron microscopy (4D-STEM) \cite{Ophus2019}, or nano-beam diffraction (NBD), with a small convergence angle $\alpha$ has recently demonstrated its power to reveal a multitude of nanoscale structural details in a very broad range of material samples, e.g. in refs \cite{Deng2019, Ozdol2015, Pekin2019}. 
Mapping the orientation of $\pi$-stacking in organic semiconductors molecular crystals was recently demonstrated under cryogenic temperatures \cite{Panova2019}, opening a new application field in beam sensitive soft materials [13]. 
Cryo-freezing the samples is one of the general strategies to slow down the structural damage and extend the dose tolerance by up to an order of magnitude before the structures break down by the incident electron beam \cite{Egerton2019, Panova2019}. 

\begin{figure}[!tb]
\centering
\includegraphics[width=8.5cm]{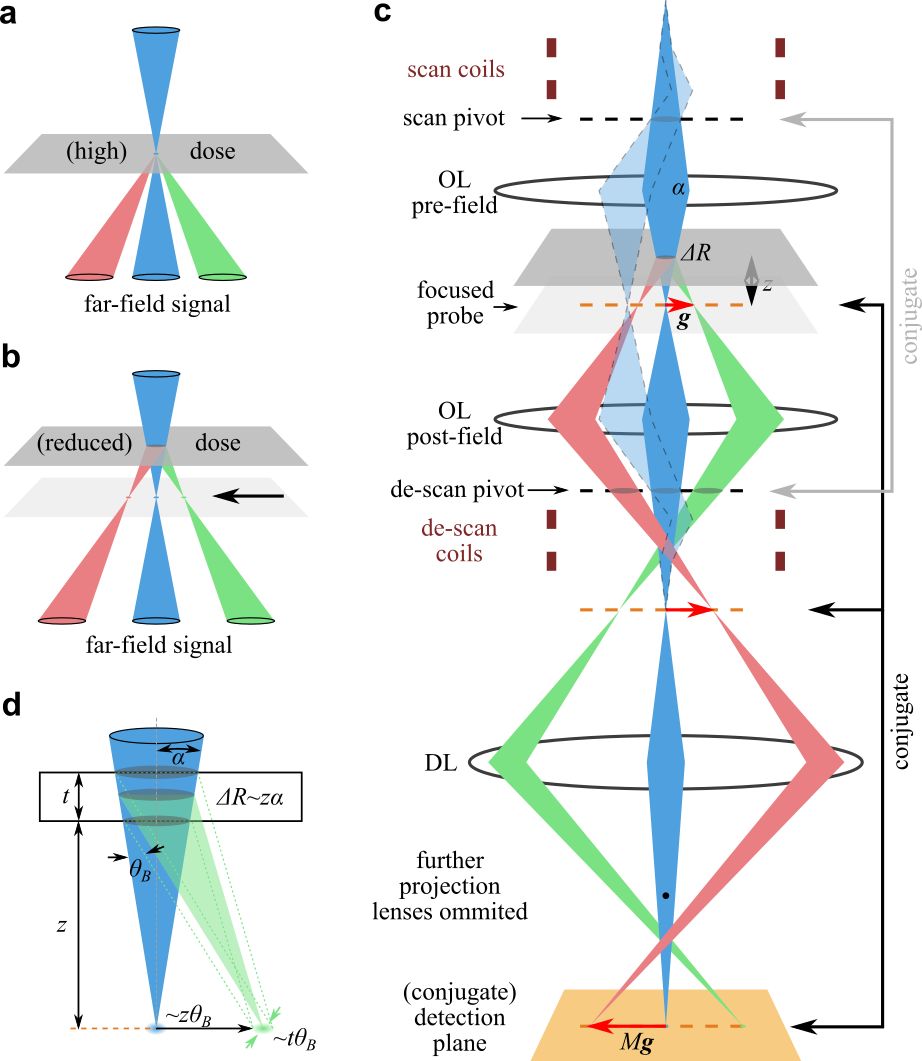}
\caption{(a) Typical STEM setup where detector(s) is(are) located at far-field. 
(b) Defocusing the probe mitigate limited dose budget for radiative sensitive samples. 
(c) Scheme of a sim-plified optic path to realize scanning confocal electron diffrac-tion. 
(d) A simple geometric consideration of spatial and an-gular resolution.
\label{fg:scheme}}
\end{figure}

Working at cryogenic temperature, however, makes {\it in situ} observation of thermal treatment induced structural evolution more difficult. 
In 4D-STEM, pixelated detectors/cameras are used to record the full 2D diffraction pattern at each probed sample position, allowing full reciprocal space details of the scattered intensities to be analyzed afterwards. 
Radical developments have pushed the limits of detection efficiency and camera speed \cite{Bustillo2021}, as well as ever growing computational and software algorithms. 
The standard NBD setup available on most TEMs, is however, neither optimized for dose efficiency nor for angular resolution. 
This is because a focused probe is interacting with a small sample region (high dose under a given probe current) and far field diffraction disks spread the signal over many pixels of the detector/camera (Fig.~\ref{fg:scheme}a). 
In applications of diffraction imaging, it is the position of Bragg reflections, their summed intensities, and their in-plane orientations which provide rich information on the crystalline phase \cite{Deng2019}, strain state \cite{Ozdol2015, Pekin2019} and orientation \cite{Panova2019} of the underlying diffracting lattices. 
The distribution of intensity in the beam disks does not typically contain additional information, but rather spreads the already low signal to many detector pixels. 
This lowers the SNR for a given detector, complicates the post-acquisition processing \cite{Bustillo2021}, and reduces the angular resolution because of disk overlap, when study large unit cell samples such as organic crystals (typically few nanometers). 
For example, the Bragg angle for a 2 nm lattice spacing is only ~0.6 mrad using 200 keV incident electrons, comparable to the probe semiangle in many NBD experiments. 

To mitigate the challenge of our limited dose budget, we can defocus the probe at the cost of spatial resolution, or use a shorter camera length, i.e. apply angular under-sampling to enhance SNR. 
Both approaches however would not improve the angular resolution of the diffraction patterns. 
Reducing the convergence angle is another possible solution, which could be achieved by using (1) condenser zoom, (2) customized small probe-defining apertures as recently explored \cite{Bustillo2021}, or (3) an objective lens with longer focal length, e.g. condenser mini lens in low-mag (LM-)STEM or Lorentz STEM \cite{McVitie2015}.
Condenser zoom can lower the convergence semi-angle to 0.5--1 mrad \cite{Wu2017}, which can still cause disk overlap for lattice planes of \texttildelow 1.2--2.4~nm distances using a 200 keV primary beam. 
Customized apertures require replacement of the standard apertures, reducing the flexibility of the instrument. 
In addition, cutting down the convergence angle with a smaller aperture by a factor of $n$ lead to lowering of current by a factor of $n^2$ which may result in a very low beam current which can make the experiments challenging to perform. 
Using a long focal length for the objective lenses suffers from very large aberrations \cite{McVitie2015}, as probe correction of these weak lens are not routinely available. 
Furthermore, using a weak objective lens means a long camera length, causing diffraction patterns that produce signals beyond the camera/detector field of view. 
Dedicated alignment of the projection system may shrink the effective camera length. 
However, medium to high angle scattering would still be blocked by the differential pump aperture in the projection chamber. 
Therefore, for diffraction imaging studies of soft materials, we want to develop alternative flexible methods, optimized for both dose efficiency and angular resolution, while retaining sufficient spatial resolution.

Here, we introduce a new diffraction imaging modality which, in contrast to 4D-STEM, uses the imaging mode of the microscope rather than the diffraction mode. 
This confocal optical setting \cite{Frigo2002, Wang2011}, albeit differing from typical scanning confocal electron microscopy (SCEM), is combined with a small convergence angle $\alpha$ and a large sample defocus z to obtain sharp diffraction spots. 
We call this technique 4D scanning confocal electron diffraction (4D-SCED) and use it to study the structure of molecular nano-crystallites in BHJ thin films composed of DRCN5T:PC71BM. 
We show that the 4D-SCED method has (1) high angular resolution for investigating the rich structural information of the molecular crystals, and (2) it can reduce dose by about an order of magnitude compared to the standard NBD setup used in many 4D-STEM applications. 
We further demonstrate that 4D-SCED even enables in situ monitoring of structural evolution and growth of nano-crystallites at elevated temperatures. 

In a diffraction limited STEM setup, where small con-vergence angles are applied, and lens aberration can be well omitted, spatial and angular resolution is well described by the Abbe equation: 
\begin{equation}
d \approx \frac{\lambda}{2{\rm sin}\alpha}
\label{eq:Abbe}
\end{equation}
In a typical NBD setup, the sample plane is coincident with plane of probe with sharply focused probe of size d at the left side of the equation; and detector is at far field, or conjugate plane of the aperture, which defines $\alpha$, thus disk patterns are detected (Fig. 1a). 
Due to the reciprocal relationship, one can optimize spatial or angular resolution on either side of the equation for a given incident electron wavelength. 
In view of the reciprocal relationship, exchanging the planes of probing and detection would allow for an extended illumination area (reduced dose) and sharply focused diffraction spots (high SNR and angular resolution) on a standard instrument without customized apertures. 

Examining the beam path in the case of a defocused probe (obtained by raising the sample, Fig.~\ref{fg:scheme}b), the cross-over of the direct and Bragg-diffracted beams are generated between the sample and the far-field diffraction pattern. 
Detection of these cross-over points can be easily realized using a confocal optics setting (Fig.~\ref{fg:scheme}c). 
In this setup, the detector plane is set to the confocal plane which is conjugate to the cross-over pattern (focal plane), i.e., the diffraction/intermediate lens is working in an imaging mode. 
Focused diffraction information is then formed at the stable confocal plane. 
Lowering the sample relative to the beam focus can also create a diffraction pattern at the confocal plane, but the pattern is rotated by 180$^{\circ}$. 
Scanning (shifting) the probe over the sample creates a 2D real space image grid, therefore de-scan of the probe after the imaging lens is required to stabilize the image of probe (and thus also the diffraction patterns) on the detection plane (Fig.~\ref{fg:scheme}c). 
This requirement is typical for scanning confocal electron microscopy (SCEM) \cite{Frigo2002}. 
Since it is the diffraction information which is of interest, and the full diffraction pattern is recorded in 4D datasets, we call this setup 4D scanning confocal electron diffraction (4D-SCED). 
A similar optical setting which can obtain spot diffraction patterns in TEM (without scanning, and thus not spatially resolved) was proposed by Midgley \cite{Midgley1999} and Morniroli et al. \cite{Morniroli2008}. 
In both cases, spatial resolution and dose efficiency was not considered, and a large convergence angle (up to a few degrees) and large defocus (tens of $\mu$m to mm) were applied. 4D-SCED in contrast uses a small convergence angle and appropriate defocus value to balance spatial and angular resolution, as will be discussed in the following.

We use a simple geometric model (Fig.~\ref{fg:scheme}d) to examine the experimental parameters and discuss the achievable spatial and angular resolution. 
As illustrated, the diffraction ``spots'' in SCED are not identical to that of real far field pattern, e.g., selected area electron diffraction (SAED) using parallel illumination, due to the intrinsic $z$-sensitivity of scattering signals in confocal setup \cite{Zheng2016}. 
The separation of the diffraction information, $g\sim z\theta_B$ ($\theta_B$ is the Bragg angle), is dependent on the defocus $z$, while the spread of the diffraction spots depends on the coherent length along the sample normal direction and can be estimated geometrically to be in the order of $t\theta_B$, for a sample with thickness $t$. 
At defocus values comparable to sample thickness, the complete wave-optical simulation and full dynamical beam-specimen interaction should be considered, which has been explored in detail with the goal of extracting high-resolution 3D information in SCEM, e.g. in refs \cite{Wang2011, Dwyer2012}. 
To obtain sharp spot patterns for nanoscale crystallography studies, we want to suppress the diffraction information between the primary beam and the diffraction beam, and therefore the spread of the diffraction spots should be far smaller than the separation, giving the condition $z\gg t$. 
In addition, a homogeneously thin, flat sample without tilt/bending is desirable in order to position a region of interest at the same defocus $z$. 
For a typical TEM specimen with $t<100$~nm, a defocus $z>2$~$\mu$m already results in a sharp spot diffraction pattern (cf. supporting information Fig. S1). 
For a given defocus, the spatial resolution is governed by the interaction area of the probe which is determined approximately by the convergence semi-angle $\alpha$ and defocus $z$ via (Fig.~\ref{fg:scheme}d) 
\[\Delta R \approx z\alpha	\]
Therefore, at a given defocus $z$, smaller convergence angle $\alpha$ is preferred to gain higher spatial resolution. 
For small probe convergence, the probe size $d$ at focus is limited by diffraction via the Abbe relationship given in Eq.~\ref{eq:Abbe}.
The angular resolution can then be estimated to be on the order of $d/z$ \cite{Midgley1999}. 
For a 200 keV primary beam energy ($\lambda=2.504$~pm) and a convergence semi-angle of $\alpha=1$~mrad, a probe size of \texttildelow 1.2~nm can be achieved. 
With defocus set to $z=5$~$\mu$m, a spatial resolution of \texttildelow 5~nm and angular resolution of \texttildelow 0.24~mrad would be expected. 
A more quantitative insight requires wave-optical simulation, which will be addressed in a follow up work. 
However, the ultimate achievable spatial resolution of beam-sensitive samples is typically determined by the available dose budget and detection efficiency, as well as speed of the used camera (cf. method section).

\begin{figure}[!tb]
\centering
\includegraphics[width=8.5cm]{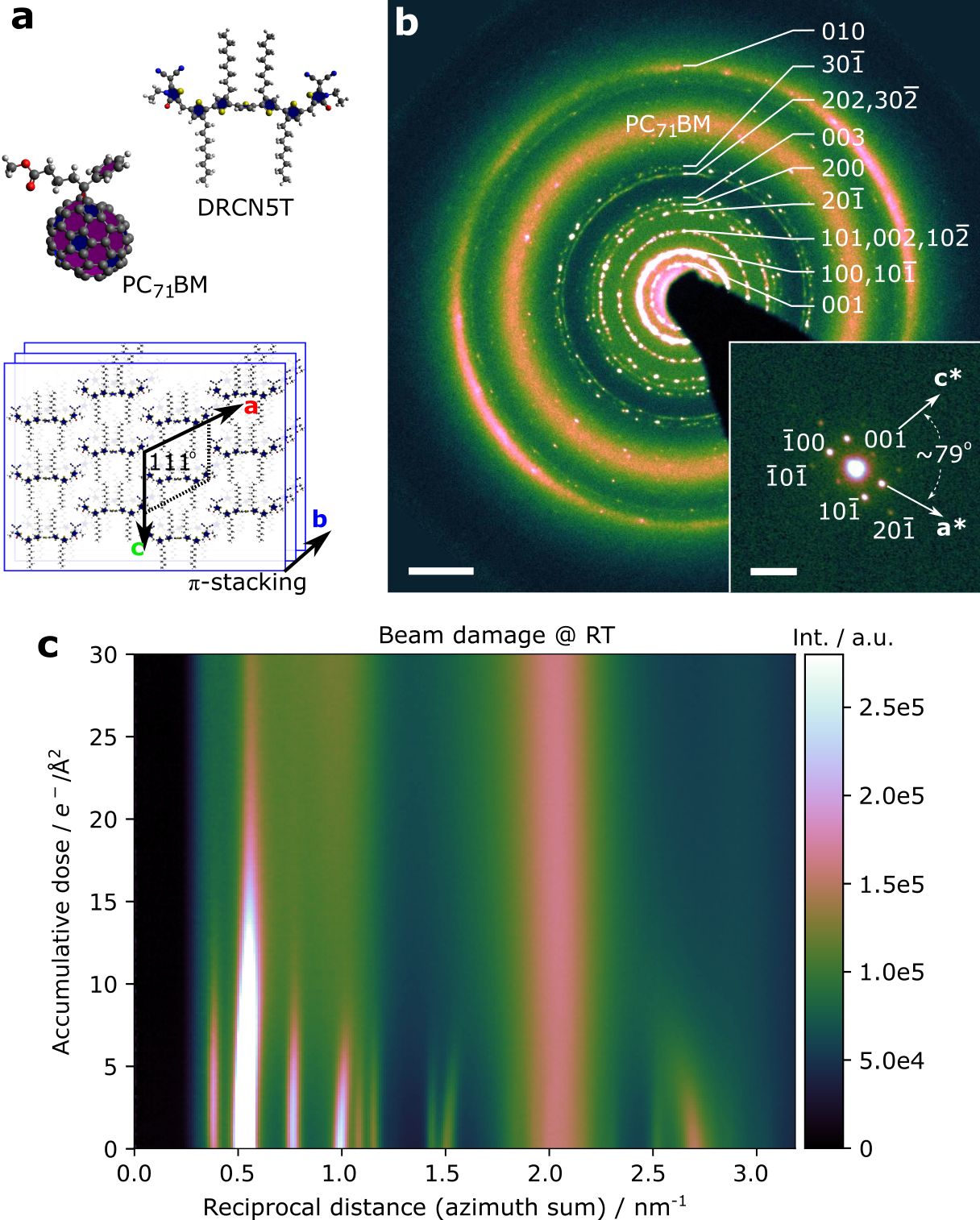}
\caption{(a) molecular and crystal structure of donner component DRCN5T and acceptor component PC$_{71}$BM. 
(b) (elastically filtered) selection area electron diffraction pattern from a sample region of 3.5 $\mu$m across. The inset shows an indexed single diffraction pattern of a face-on DRCN5T crystalline domain extracted from a SCED dataset (cf. Fig. 3). 
(c) azimuth integrated SAED as function of the accumulative dose. Scale bars: 2 mrad. Details in text.
\label{fg:damage}}
\end{figure}
\begin{figure*}[!tb]
\centering
\includegraphics[width=18cm]{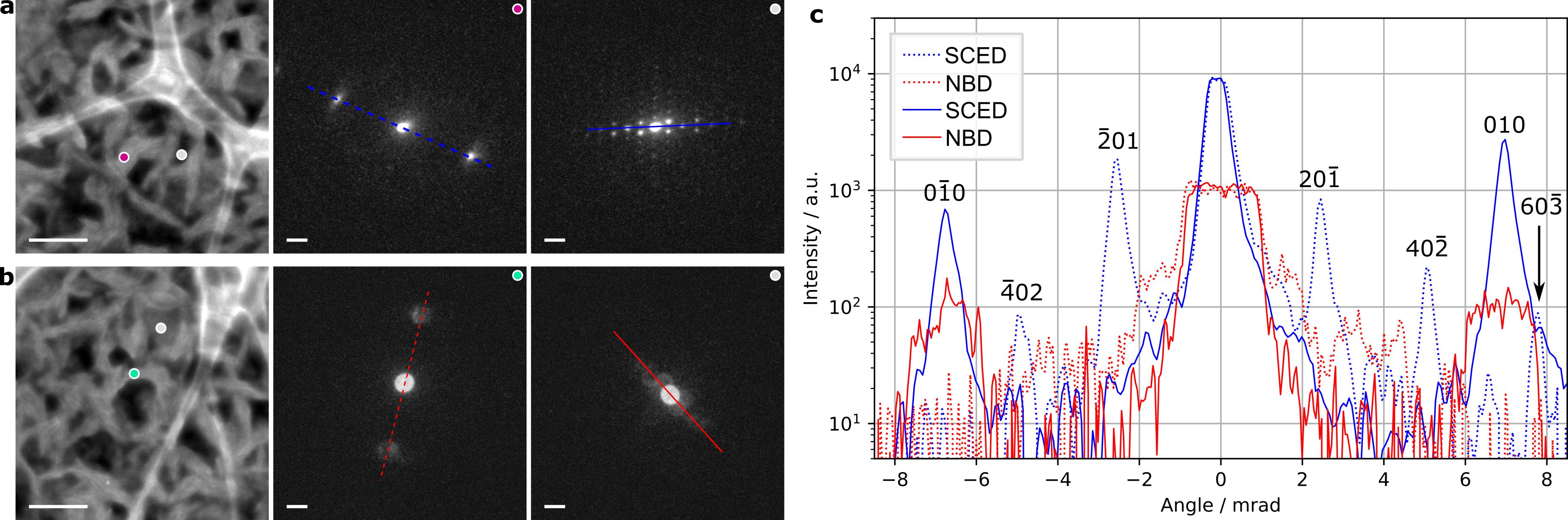}
\caption{Comparison of the raw diffraction signal intensity of SCED and NBD under identical dose conditions and spatial and angular sampling from fresh areas of a solvent vapor annealed DRCN5T:PC$_{71}$BM blend sample. 
(a) SCEM-ADF image simultaneously ob-tained during the SCED data acquisition. 
(b) STEM-ADF image simultaneously obtained during NBD data acquisition. 
Single raw diffraction patterns extracted from the marked dots in (a) and (b) representing the edge-on and face-on DRCN5T nano-crystallites respectively, are extracted and show as insets below. 
(c) Line profiles comparing the signal to noise of raw patterns represent more than an order of magnitude higher SNR, whole pattern signal analysis shown in Fig. S3.
\label{fg:SNR}}
\end{figure*}

With 4D-SCED, we have characterized the active layer of a BHJ solar cell and show the highly improved dose efficiency and angular resolution compared to 4D-STEM (or NBD) under identical illumination condition (and thus dose budget). 
For this purpose, we choose a BHJ film comprising a blend of a small molecule quinquethiophene-based electron donor DRCN5T and a fullerene acceptor PC$_{71}$BM which has been treated by solvent vapor annealing (SVA) in CS$_2$ for 840s. 
Solar cells based on this system has been shown to have high efficiency and stability \cite{Min2016}. 
Moreover, the nanomorphology and crystallinity show a clear correlation to device performance, which has been shown to depend sensitively on the processing conditions \cite{Min2017}. 
In a recent structural analysis using grazing incidence wide angle X-ray scattering (GIWAXS), the unit cell structure of the small molecule crystal DRCN5T was determined, and a coexistence of (molecular) face-on and edge-on crystalline domains were deduced \cite{Berlinghof2020}. 
The molecular orientation, particularly the $\pi$-stacking orientation, within the nanoscale percolating fiber network is the key to understand the transport properties. 
However, the important question of how the molecular orientation (which can be determined on average from other techniques such as GIWAX) locally relates to the nano-morphology of the donor-acceptor blend and how both evolve upon processing have not yet been answered.

Figure~\ref{fg:damage}a shows the chemical structure of PC$_{71}$BM and DRCN5T, respectively. 
The in-plane tiling and out-of-plane stacking of the small molecule is also schematically depicted. 
Figure~\ref{fg:damage}b present an elastically filtered SAED (cf. method) pattern of the sample acquired at room temperature (RT) with a total electron dose of \texttildelow 0.8~$e^-/{\AA}^2$. 
In the experiment, the diffraction rings fade out very rapidly. 
A time series (and thus accumulated dose) of the SAED was recorded, and the azimuth integrated profiles as function of accumulative dose is plotted in Fig.~\ref{fg:damage}c. 
While the in-plane molecule planes, i.e., $\{100\}$ diffraction ring at 0.55~nm$^{-1}$, survived beyond \texttildelow 15~$e^-/{\AA}^2$, the $\pi$-stacking, i.e., $\{010\}$ diffraction ring at 2.65~nm$^{-1}$, can only withstand a beam dose below \texttildelow 5~$e^-/{\AA}^2$. 
We note that this critical dose is comparable to that of Polyethylene and is orders of magnitude more vulnerable than metal organic frame-works, which can tolerate \texttildelow 100--1000~$e^-/{\AA}^2$, as measured in controlled experiments \cite{Egerton2019}. 
The individual diffraction rings follow different trends upon beam bombardment that reveal the time evolution of structural damage. 
It is apparent that almost all in-plane diffraction rings $\{h0l\}$ expand to higher values while the \{010\} diffraction ring shrinks, indicating that the crystal order of the $\pi$-stacking expands immediately upon beam bombardment while the in-plane order of the molecule is gradually shrinking. 
Working under cryogenic temperature helps to preserve the crystalline order of the small molecule to about 4 times the total electron dose compared to RT (Fig. S2).
However, ice formation was observed (in both SAED and real space imaging) upon illumination. 
This disturbed the analysis of the native structure of the sample. 
The SCED and NBD 4D-STEM experiments were therefore performed at RT. 

To illustrate the enhanced SNR of 4D-SCED compared to NBD 4D-STEM using standard apertures, we compare the datasets acquired from the same sample at neighboring fresh areas, under identical dose conditions. 
We emphasize here the smallest available convergence of 0.85~mrad on our instrument was applied. 
Figure~\ref{fg:SNR}a--b show the sample areas and representative raw diffraction patterns extracted from the marked regions using 4D-SCED (Fig.~\ref{fg:SNR}a) and NBD 4D-STEM (Fig.~\ref{fg:SNR}b), respectively. 
Each of the two patterns represent edge-on (color dot) and face-on crystallite (gray dot). 
In the face-on case the spots from the $\{h0l\}$ planes, separated by angles $<1$~mrad are clearly resolved in the SCED mode showing very high signal in-tensity, while they are heavily overlapping in the NBD mode. 
Apart from the disc overlap, the SNR is much inferior to that in SCED. 
With customized smaller apertures, thus smaller beam disks, more concentrated diffraction signals are expected, and the difference would be less striking. 
At same beam flux, the difference of SNR will dependent on the exact shape of probe profile and electron-sample interaction circle in the respective setups. 
The line intensity profiles from the NDB and SCED data are extracted along the blue and red lines, respectively, and are shown in Fig.~\ref{fg:SNR}c. 
The dashed lines compare data from the edge-on domains while solid lines compare the data from face-on domains. 
Obviously, almost all diffraction peaks in 4D-SCED show an order-of-magnitude higher intensity compared to NBD. 
This is most prominently revealed by the \{603\} peaks which are clearly visible in SCED, while in NDB they are buried in the noise floor. 
Comparing the SNR from the whole maps (cf. Fig. S3) reveals an average enhancement of the peak intensity by an order-of-magnitude in 4D-SCED. 
Since the noise level (\texttildelow 10 counts in current case) is intrinsic to the camera, the SNR level in SCED is therefore about an order-of-magnitude higher than NDB under the applied acquisition conditions. 
This means under same detectability criteria of our camera, SCED requires much less dose for Bragg peak detection, corresponding to higher dose efficiency. 

\begin{figure*}[!tb]
\centering
\includegraphics[width=18cm]{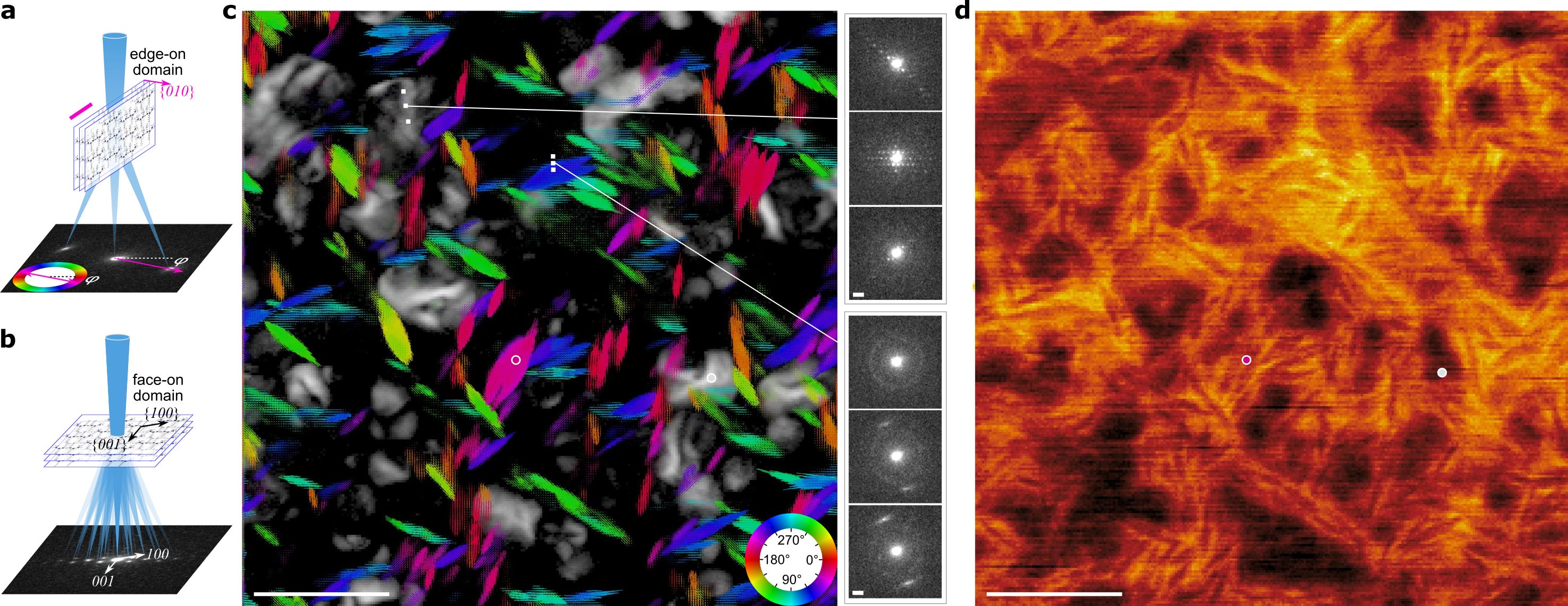}
\caption{Visualization of the orientation of donor nano-crystallites and distribution of the PC$_{71}$BM acceptor. 
Scheme (a) uses a color wheel method to encode the edge-on domain orientation at a probed location; (b) demonstrates our ability to determine the grain orientation of the face-on domains. 
(c) Visualization of the whole scanned field of view. 
Insets on the right are raw dif-fraction patterns extracted from the white dots. 
With virtual annual aperture including only the angular range of PC$_{71}$BM in each diffraction pattern, the location of the acceptor is shown in (d). 
Scale bars: 500~nm in (a) and (c); 2~mrad in insets of (c).
\label{fg:visualize}}
\end{figure*}

With its high angular resolution, the patterns acquired via SCED enables mapping of the orientation of nano-crystallites not only in edge-on (large diffraction angles) but also in face-on orientation (small diffraction angles). 
Furthermore, the sharp diffraction spots in SCED can be used to locally analyze the crystallographic structure of individual nano-crystallites, which is more difficult in the comparable NBD experiments. 
To demonstrate this, the inset in Fig.~\ref{fg:damage}b shows an indexed SCED pattern extracted from a face-on domain, which agrees well to that obtained from GIWAX studies \cite{Berlinghof2020}. 
Since the charge carrier transport properties of molecular crystals are highly anisotropic, the orientation of the molecular crystals in the domains and the micrometer scale percolation of the domains are critical to pinpoint the device performance. 
Figure~\ref{fg:visualize} visualizes the orientation of edge-on crystal domains using the color wheel method, and the location of the face-on domains are displayed as superimposed grayscale maps.
The short color segments represent the backbone of DRCN5T molecules, which is perpendicular to the \{010\} diffraction g-vectors (schematically shown in Fig.~\ref{fg:visualize}a). 
It is now clear that the $\pi$-stacking of the DRCN5T molecules is aligned along the long axes of the nanoscale fiber structures, and the $\pi$-stacking plane normal, i.e., $\langle 010\rangle$ direction, along the short axis of fibers. 
In this sample (SVA treated in CS$_2$ for 840s), the face-on domains cover roughly circular areas ranging from a few tens up to a several hundred nanometers in size. 
Within a face-on domain, the molecular crystals show tilted and twisted diffraction patterns (top-right insets in Fig.~\ref{fg:visualize}c) indicating either a strong bending of the domain or the existence of sub-grains and grain boundaries. 
Furthermore, successive raw diffraction patterns across the interface of edge-on domains reveal that PC$_{71}$BM is enriched mainly at the donor interface. 
This is best visualized in the (virtual) annual dark field image using the diffraction information characteristic to PC$_{71}$BM, i.e., between 1.8 and 2.3~nm$^{-1}$, in Fig.~\ref{fg:visualize}d. 
We note that mapping the PC$_{71}$BM was not feasible using a simple virtual aperture method due to the much inferior angular resolution and overlapping diffraction information. 
In samples which have undergone SVA using CHCl$_3$ as the solvent, similar fiber-like edge-on domains were observed but the face-on domains also resemble a fiber-like shape (cf. Fig. S5). 
Considering that the DRCN5T molecule has a flat $\pi$-conjugate plane and its flat tiling into crystalline grains, electron wavefunction overlapping along the molecule out-of-plane direction, i.e., the molecular crystal [010] direction, is not likely [16]. 
The high-mobility direction of charge carrier transfer is dominated by pathways in the crystal a-c plane, which is determined to coincident with the long axes of the fiber. 

\begin{figure*}[!tb]
\centering
\includegraphics[width=15cm]{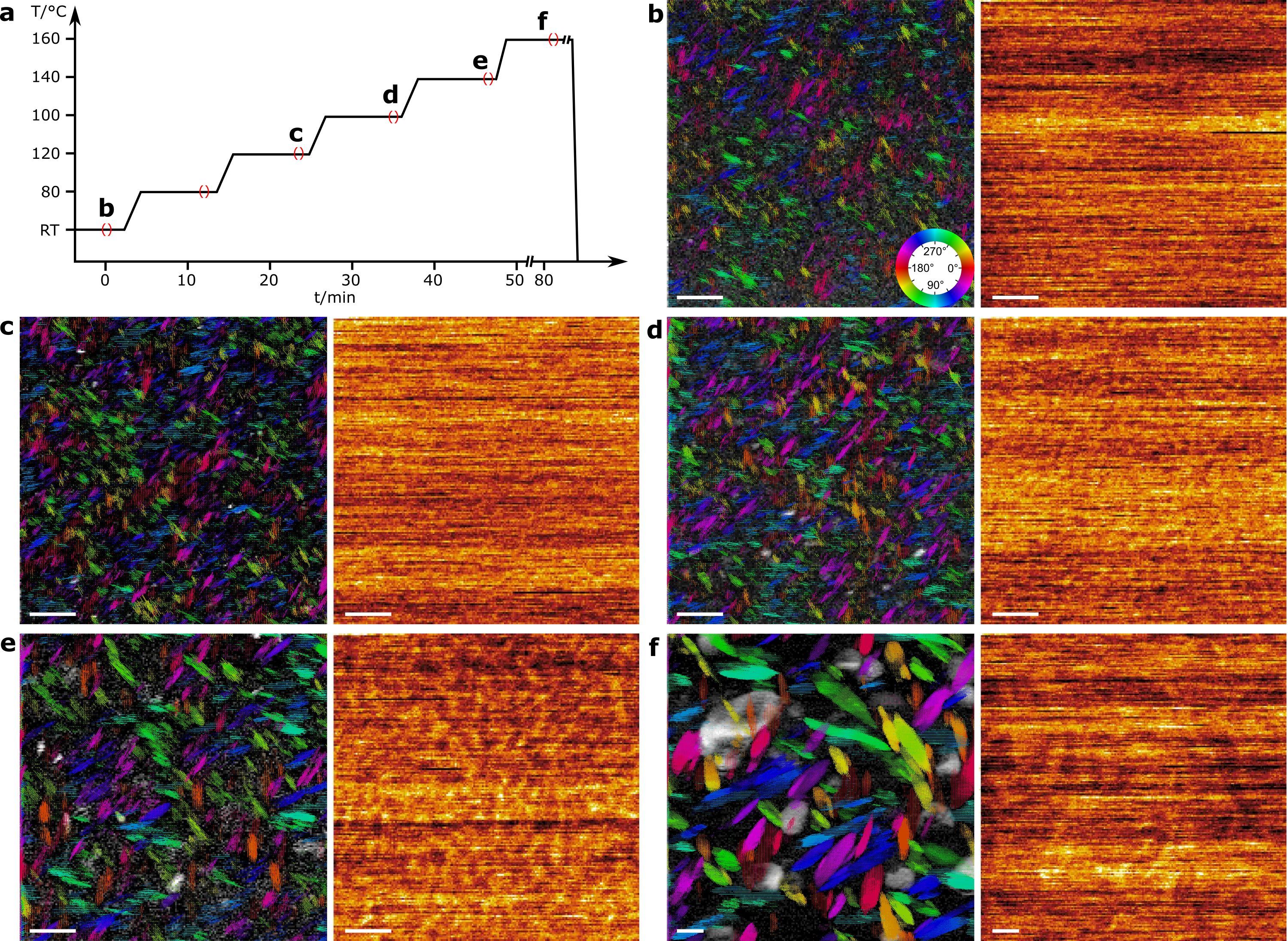}
\caption{Observing the structural evolution upon thermal annealing of DRCN5T:PCBM thin film in the TEM. (a) Temperature profile applied in the experiment. The red brackets with (b) – (f) indicate the time (of about 2 minutes each) spent on the SCED data acuisition. The data at each stage is visualized in (b)–-(f) with the same scheme used in Fig.~\ref{fg:visualize}. Scale bars: 200 nm.
\label{fg:insitu}}
\end{figure*}

Finally, we use 4D-SCED to study the structural evolution of DRCN5T:PC$_{71}$BM during annealing at elevated temperatures by in situ heating the sample thin film in the vacuum of TEM. 
Figure~\ref{fg:insitu}a show the temperature profile applied during the experiment. 
4D-SCED datasets were recorded sequentially with only few seconds between scans, to move the sample to a nearby fresh (un-illuminated) region. 
Figure~\ref{fg:insitu}b--f show visualizations of the sample morphology and orientation at different thermal annealing steps. 
The as-casted samples show already a certain degree of $\pi$-stacking ordering, but barely any face-on domains. 
Very small ($<50$~nm long) edge-on domains were visible, having a rod-like shape in projection. 
The acceptor component PC$_{71}$BM appears homogeneous. 
After the sample have been annealed under 100$^{\circ}$C for $\sim8$~min, a few tiny face-on domains spanning only 3--4 probed pixels (15–-20~nm) start to emerge. 
Here, the growth of the edge-on domains in size is not yet obvious. 
After raising the temperature to 120$^{\circ}$C and holding for about 8~min, growth of both edge-on and face on domains were observed, which is accompanied by phase separation of the PC$_{71}$BM. 
Raising the temperature again to 140$^{\circ}$C and holding 8 min, the edge-on domains become sharper and enrichment of PC$_{71}$BM towards the edge-on DRCN5T domains were observed. 
Finally, after the sample has been held at 160$^{\circ}$C for about 10 mins, the edge-on domains grow to an apparent length of 300-–500 nm and large face-on domains also appear. 
Further annealing does not result in additional growth, which is likely due to the completed phase separation and depletion of the small molecules. 
Interestingly, during the growth of fiber-like edge-on crystalline domains, the aspect ratio about 3--4 seem to remain throughout the entire thermal annealing processing. 
With the observed crystallographic orientation relationship to the fiber morphology, this indicates the growth of the nano-crystallites must be faster in the crystallographic $a$-$c$ plane, and much slower expanding along the $\pi$-stacking $\langle 010\rangle$ direction. 
This further hint that the growth (attachment of new molecules to crystalline seeds) via out-of-plane $\pi$-electron interaction is weaker than the in-plane interaction in this materials system.

In conclusion, we have demonstrated 4D-SCED as a highly dose-efficient and high angular resolution diffraction imaging method. Under optimized acquisition parameters, we observed the annealing-induced growth and structural evolution of nano-crystallites at sub-5 nm spatial resolution, under a dose budget of \texttildelow 5~$e^-/{\AA}^2$. 
When coupled to state-of-the-art high DQE detectors and more advanced data mining algorithms to extract subtle signals out of the large datasets like those used in NBD studies, we expect further improvements in the dose efficiency. 
We believe this new 4D-STEM experimental setup will open new possibilities in the study of nano-crystallography of soft materials.

\begin{small}
\section*{Materials and methods}
{\it Sample preparation.}

The bulk heterojunction thin films were fabricated by a spin-coating process on ITO-coated (thickness: 350 nm to 400 nm) glass substrates ($1.0\times 1.0$~in$^2$, Hans Weidner GmbH, N\"urnberg Germany). 
Pre-structured ITO coated glass substrates were subsequently cleaned in acetone and isopropyl alcohol for 10~min each.
After drying, the substrates were bladed with 40~nm poly(3,4-ethylenedioxythiophene) polystyrene sulfonate (PEDOT:PSS, Clevios P VP Al 4083, Heraeus, Hanau, Germany). 
The DRCN5T bulk films were spin-coated (1500~rpm) under inert gas atmosphere from the solutions of DRCN5T (purity $\geq 99$\%, 1-Material Inc., Dorval, Canada) and PC$_{71}$BM (purity $\geq 99$\%, Solenne B.V., Groningen, Netherlands) (1:0.8~wt.\%) in chloroform leading to a film thickness of about 80~nm as estimated by a profilometer. 
The DRCN5T and PC$_{71}$BM solutions were stirred at 40$^{\circ}$C and 150~rpm before mixing and spin-coating. 
For the SVA post-processing procedure, the samples were loaded in the middle of a closed Petri dish containing 120~$\mu$l of carbon disulfide (CS$_2$) solvent.

{\it Electron Microscopy.}

Energy filtered selected area electron diffraction experiments were performed on a ThermoFisher Scientific (TFS) monochromated, double C$_s$ corrected Titan Themis microscope operating at 300~kV, and NBD 4D-STEM and 4D-SCED on a TFS probe corrected Spectra microscope equiped with a X-CFEG gun and operated at 200~kV. 
The Themis is equiped with a regular Ceta (Scintilator coupled CMOS) camera capable to run at a maximum frame rate of ~35~fps @ $512\times512$~pixels with dynamic range of about 13-bit (operated via TIA software), and the Spectra is coupled to a high sensitivity Ceta-S camera with maximum speed of ~300~fps @ $512\times512$~pixels with dynamic range of about 12-bit (operated via Velox). 
For NBD, the optics were set to micro-probe STEM mode, with the smallest standard 50~$\mu$m C2 aperture and at the limit of convergence zoom. 
The smallest achievable convergence semi-angle are {\em calibrated} to be $\alpha=0.7$~mrad (on the Themis) and $\alpha=0.85$~mrad (on the Spectra), respectively. 
On both microscopes, we noticed the factory alignment displays wrong convergence angle almost half of our calibrated values. 
To reach the confocal diffraction condition, basic column alignment of micro-probe STEM (at the smallest $\alpha$) were performed first, then diffraction lens was switched to imaging mode. 
Image magnification is first set to a high value, $\sim$100~kx, to confirm that the confocal plane is properly reached, which is done by going through focus of the probe-defining (i.e, C3) lens and find the minimum size of the probe. 
After this point, defocusing is achieved only done via moving sample up and down. 
The magnification of diffraction pattern will change upon shifting the sample, give full flexibility to tune/balance of the desired spatial and angular resolution. 
For a given defocus (i.e., sample offset) value, the image magnification is set so to appropriately cover the range of diffraction vectors of interest on the camera. 
The scan and de-scan pivot points were carefully aligned. 
Finally the shape of probe is determined by the aberrations of the probe-forming lens, and distortions of the diffraction pattern can be corrected by carefully aligning the objective imaging lens (cf. Fig. S7).

{\it 4D datasets acquisition due to dose limited resolution.} 

In all experiments, cameras were operated at their highest speed and other acquisition parameters (probe current, dwell time, defocus and scanning pixel distance) were estimated to keep both total and instaneous dose budget well below $10~e^-/{\AA}^2$. 
To estimate the dose limited resolution in a scanning probe experiment, we can use instaneous dose at any probing point
\[
D\left[\frac{e^-}{\AA^2}\right] = \frac{I_p\cdot 10^{-12}\left[C/s \right]\cdot 6.24 \cdot 10^{18} \left[e^- \right] \cdot t_{px} \cdot 10^{-12}\left[ s\right]}{\pi\cdot \Delta R^2 \left[ \AA^2 \right]}
\]
with $I_p$ probe current (in pA), $t_{px}$ dwell time at each pixel (in ms) and $D$ dose budget (in $e^-/\AA^2$) that can be evaluated from SAED experiments, to be
\[
\Delta R \left[ \AA \right] \approx 45 \sqrt{I_p \left[pA \right] \cdot t_{px} \left[ ms \right] / D \left[ e^-/\AA^2 \right]}.
\]
Considering the experiments performed on the Titan Themis platform, inserting the fastest camera frame time (roughly equals to the dwell time) $t_{px}=13.5$~ms and dose bugest $D=5~e^-/\AA^2$, a dose limited resolution of $\Delta R\approx74~{\AA}$ is obtainable when setting probe current to $I_p=1$~pA. 
With the smallest available convergence angle on our microscope under mirco-probe mode $\alpha=0.7$~mrad, a defocus value $z>\Delta R/\alpha\approx10~\mu$m is needed. 
On the Spectra platform (smallest $\alpha=0.85$~mrad) at the same dose budget of $D=5~e^-/\AA^2$ and probe current of $I_p=1$~pA, with shorter dwell time $t_{px}=3.5$~ms, spatial resolution $\Delta R\approx38$~{\AA} is possible. 
To achieve this condition, a defocus value $z>\Delta R/\alpha\approx 4.5~\mu$m is required. 
The scan pixel distance (sampling size) is set to slightly larger than the abovementioned resolution limit. 
For {\em in situ} experiments, it is important to balance spatial resolution required to reveal small structures in the early state, field of view to cover statistical relevant sample region, reasonable short time of data acquisition to minimize image distortion due to thermal drift. 
We took the following experimental parameters: $180\times 180$ grid area with 4~nm STEM probe step size, defocus of 2.5~${\mu}$m and probe current below the measurable quantity of 1~pA in the first few frames, and STEM probe step size increased to $\sim$6~nm when large structured emerged and field of view was limited.

{\it Data handling, pre-processing and visualization. }

The acquired data were pre-processed in Gatan DigitalMicrograph software with public plugins and homemade scripts. 
The acquired data is firstly aligned to account for the shift of center beam. 
Elliptical distortion of the sum diffraction pattern was observed in some of the SCED experiments, which can be hardware corrected using the objective lens stigmator (cf. Fig. S6). 
The distortion was observed to not depend on the scanning position when the scanning field is within few micrometers.
This makes the post-acquisition software correction of individual pattern feasible based on the distortion evaluated from the summed patterns. 
For this, we applied the evaluation and correction method developed by Mitchell and Van den Berg \cite{Mitchell2016}. 
Due to the high SNR in the SCED datasets, the diffraction signals at small angles were obvious and straightforward for subsequent analysis using simple virtual apertures; and no hardware elastic filtering (like in Fig.\ref{fg:damage}b) or post-processing background subtraction \cite{Panova2016} were necessary. 
The 2D orientation map of edge-on crystal domains was obtained by directly evaluating the virtual dark field image intensities using sector apertures of radius between 2.6 and 2.8~nm$^{-1}$ and angular increment of $n$ degrees. 
In this way a total of $360/n$ dark-field images were calculated, and each image corresponds to in-plane diffraction angle $\phi=ni$ defined by the sequence $i$ of the virtual apertures. 
This reduces the 4D dataset to a 3D dark field image data cube. 
At any (real space) scanning pixels, we search for the maximum intensity $M$ along the dark field sequence and mark its location $\phi$, which together defines 2D grid of complex numbers $M(x,y)e^{i\phi(x,y)}$ (equivalent to 2D vectors).
For face-on crystallites or overlapping crystallites, multiple diffraction spot may appear at same diffraction angle $q$, and the above algorithm fails to capture same orientation crystal domains with slight out-of-plane tilt (which result in change of the center of Lauer circle and flicking of diffraction spots intensities). 
For this, we apply a rotation invariant template matching method to find location of face-on domains and determine its rotation angle. 
Here, the rotation invariant template matching is converted to a shift invariant template matching via cross-correlation of the polar transformed diffraction patterns. 
Finally, the processed data in form of 2D vectors were visualized using Python with Matplotlib, OpenCV and numpy.

\section*{Author contributions}

MW and ES conceived the idea and designed the experiments. 
CH prepared the samples and performed extensive TEM characterization of the samples. 
MW and CH performed the in situ experiments. 
MW developed 4D-SCED technically, and implemented algorithms to analyze the data, drafted the manuscript with input from all authors. 
CO examined the 4D datasets to verify the results using alternative analysis and visualization routines. 
All authors have given approval to the final version of the manuscript.

\section*{Acknowledgments}
We acknowledge financial support from Deutsche For-schungsgemeinschaft (DFG) via the research training school GRK 1896: ``In-Situ Microscopy with Electrons, X-rays and Scanning Probes'' and the Cluster of Excellence EXC 315 ``Engineering of Advanced Materials''. 
SFB 953 ``Syntetic Carbon Allotropes''. 
Work at the Molecular Foundry was supported by the Office of Science, Office of Basic Energy Sciences, of the U.S. Department of Energy under Contract No. DE-AC02-05CH11231.

\section*{Abbreviations}
DRCN5T: 2,2'-[(3,3''',3'''',4'-tetraoctyl[2,2':5',2'':5'',2''':5''',2''''-quinquethiophene]-5,5''''-diyl)bis[(Z)-methylidyne(3-ethyl-4-oxo-5,2-thiazolidinediylidene)]]bis-propanedinitrile 

PC$_{71}$BM: Phenyl-C71-butyric acid methyl ester

\section*{Supporting Information}
Supporting figures S1-S6. 

\end{small}

\bibliography{library}

\end{document}